\documentclass[twocolumn]{emulateapj}

\newcommand{\Msun}{$M_{\odot}$}
\newcommand{\Rsun}{$R_{\odot}$}
\newcommand{\Mdot}{$\dot{M}$}

\shorttitle{Stars, accretion shocks, and dust sublimation walls in TTS}
\shortauthors{McClure et al.}

\begin{document}

\title{Curved walls: grain growth, settling, and composition patterns in T Tauri disk dust sublimation fronts}


\author{M. K. McClure\altaffilmark{1,7,8}, P. D'Alessio\altaffilmark{2}, N. Calvet\altaffilmark{1},  C. Espaillat\altaffilmark{3,9}, L. Hartmann\altaffilmark{1}, B. Sargent\altaffilmark{4}, D. M. Watson\altaffilmark{5}, L. Ingleby\altaffilmark{1},J. Hern\'andez\altaffilmark{6}}

\altaffiltext{1}{Department of Astronomy, The University of Michigan, 500 Church St., 830 Dennison Bldg., Ann Arbor, MI 48109; melisma@umich.edu, ncalvet@umich.edu, lhartm@umich.edu, lingleby@umich.edu}
\altaffiltext{2}{Centro de Radioastronom\'{i}a y Astrof\'{i}sica, Universidad Nacional Aut\'{o}noma de M\'{e}xico, 58089 Morelia, Michoac\'{a}n, M\'{e}xico; p.dalessio@astrosmo.unam.mx}
\altaffiltext{3}{Center for Astrophysics, 60 Garden Street, Cambridge, MA 02138, USA; cespaillat@cfa.harvard.edu}
\altaffiltext{4}{Center for Imaging Science and Laboratory for Multiwavelength Astrophysics, Rochester Institute of Technology, 54 Lomb Memorial Drive, Rochester, NY 14623, USA; baspci@rit.edu}
\altaffiltext{5}{Department	of Physics and Astronomy, University of Rochester, Rochester, NY 14627, USA; dmw@pas.rochester.edu}
\altaffiltext{6}{Centro de Investigaciones de Astronom\'ia (CIDA), M\'erida 5101-A, Venezuela; hernandj@cida.ve}
\altaffiltext{7}{NSF Graduate Research Fellow}
\altaffiltext{8}{Visiting Astronomer at the Infrared Telescope Facility, which is operated by the University of Hawaii under Cooperative Agreement no. 
NNX-08AE38A with the National Aeronautics and Space Administration, Science Mission Directorate, Planetary Astronomy Program.}
\altaffiltext{9}{NASA Sagan Exoplanet Fellow}

\begin{abstract}

The dust sublimation walls of disks around T Tauri stars represent a directly observable cross-section through the disk atmosphere and midplane.   Their emission properties can probe the grain size distribution and composition of the innermost regions of the disk, where terrestrial planets form.  Here we calculate the inner dust sublimation wall properties for four classical T Tauri stars with a narrow range of spectral types and inclination angles and a wide range of mass accretion rates to determine the extent to which the walls are radially curved.  Best-fits to the near- and mid-IR excesses are found for curved, 2-layer walls in which the lower layer contains larger, hotter, amorphous pyroxene grains with Mg/(Mg+Fe)=0.6 and the upper layer contains submicron, cooler, mixed amorphous olivine and forsterite grains.  As the mass accretion rates decrease from 10$^{-8}$ to 10$^{-10}$ \Msun/yr, the maximum grain size in the lower layer decreases from $\sim$3 to 0.5 $\mu$m.  We attribute this to a decrease in fragmentation and turbulent support for micron-sized grains with decreasing viscous heating.  The atmosphere of these disks is depleted of dust with dust-gas mass ratios 1$\times$10$^{-4}$ of the ISM value, while the midplane is enhanced to 8 times the ISM value. For all accretion rates, the wall contributes at least half of the flux in the optically thin 10 $\mu$m silicate feature.  Finally, we find evidence for an iron gradient in the disk, suggestive of that found in our solar system.

\end{abstract}


\section{Introduction}
\label{intro}

The excess emission seen in T Tauri stars (TTS) from 1 to 5 $\mu$m is now commonly attributed to the sharp inner edge, or `wall', where the dusty circumstellar disk reaches temperatures high enough to sublimate the dust \citep{natta+01}.  If the shape and absolute flux of the excess are known, it is possible to fit the excess with models to determine the geometry and composition of the wall \citep{muzerolle03,ajay+08,espaillat+10}.  However, the precise shape of the wall is an outstanding problem in the field.

In its original conception, the dust sublimation wall was treated as a vertical slab at a fixed temperature with uniform dust properties \citep{natta+01,dalessio+06}.  However, there are at least three major effects that may act to modify the shape.  First, the value of the sublimation temperature is dependent on the pressure of the gas in which the grains are suspended \citep{pollack+94,gail+99}.  Therefore dust in the midplane, which has typical pressures of $\sim$10$^3$ dyn cm$^{-2}$, sublimates at a higher temperature than dust in the upper layers of the disk, where pressures are typically 10$^{-6}$ to 10$^{-3}$ dyn cm$^{-2}$. The second effect takes into account the expected vertical gradient in the maximum grain size.  Large grains are predicted to settle to the midplane, leaving only submicron sized grains in the upper layers \citep{dd04}.  Smaller grains are more efficient absorbers of radiation at stellar frequencies than are larger grains of the same dust species; therefore, they achieve their sublimation temperature at a larger radial distance from the star.  The net result of these effects is that the wall has a radial spread and is closest to the star at the midplane \citep{isellanatta05,ajay+07,nagel+13}.  

The third effect, which has yet to be explored in great detail, is that different dust species sublimate at different temperatures, for a given pressure \citep{pollack+94,gail+99}.  More refractory minerals, like graphite, alumina, and calcium-rich silicates, may exist interior to radii of `typical' dust sublimation temperatures, i.e. 1400 K \citep{posch+07}.  Recent interferometry work in Herbig AeBe stars has found evidence suggesting that some fraction of the NIR emission originates in hot material inside the 1400 K radius \citep[][and references therein]{ajay+08}.  It has been suggested that this emission might be optically thick gas in the inner, dust-free zone \citep{eisner+07, fischer+11}.  However, these models make assumptions about the dust composition, grain size distribution, and density structure of the inner disk, all of which influence the temperature and location at which the dust sublimates.  \citet{najita+09} find that the hot material interior to the dust sublimation wall does not posses signatures typical of the predicted gas species; instead they suggest that it might be highly refractory dust species.  

At the same time, the height of the wall relative to the disk behind it could determine how much stellar emission is incident on the outer disk; if the wall `shadows' the disk it would prevent the terrestrial planet-forming region from being heated effectively, producing less flaring \citep{natta+01,dullemond+01, meeus+01,dd04}, and potentially affecting chemical reactions that require a strong UV radiation field.  For these reasons, it is important to understand the shape of the wall and its interplay with the disk behind it.

Analysis of the wall geometry is complicated by the reliability of the measured excess, which requires an understanding of the underlying stellar and accretion properties, and knowledge of the disk properties behind the wall.  In \citet{mcclure+13} (Paper I), we presented a parametric analysis of the flux-calibrated near-infrared excesses of eight accreting TTS.  For most of these systems, the excess could be fit by two blackbodies, one at the temperature appropriate for an emitting accretion shock ($\sim$8000 K) and another at a temperature consistent with dust sublimation ($\sim$1700 K).  For a subset of targets, a third blackbody at a temperature of $\sim$800 K was required to fit the 4-5$\mu$m excess.  A simplistic estimate of the dust sublimation radii based on these temperatures suggested that the dust sublimates at 0.03 to 0.1 AU in these systems.  However, the wall solid angle required to fit the absolute flux of these excesses was large, implying wall heights of $\sim$10-30 times the gas pressure scale height, $H$.  In this work, we use the \citet{dalessio+06} wall and disk models instead of blackbodies to fit the NIR excess, allowing us to study the connection between the wall and disk.

\section{Sample and data}

From the initial set of eight CTTS in Paper I, the cool NIR excess in V836 Tau, GO Tau, BP Tau, and DE Tau was well represented by a single temperature black body around 1600 K, while FN Tau, DS Tau, CI Tau, and DR Tau were best fit by two blackbodies at 1700 and 800 K.   The relatively narrow range of spectral types in this sample (K7-M1.5) eliminates one source of variation in our modeling attempts.  However, we further restrict our sample here to systems with a well-characterized inclination angle between 40 and 70\degr, i.e. V836 Tau, GO Tau, DE Tau, and CI Tau.  As we describe in \S \ref{dwall}, our wall models approximate curvature with two layers, each with a vertical face.  The maximum flux for a vertical wall will be achieved at inclinations $\sim$60\degr, and the approximations we employ for the curved wall break down for disks with a more face-on inclination (e.g. FN Tau, BP Tau). DS Tau did not have a reliable inclination, while we excluded DR Tau because its total luminosity is dominated by its accretion shock rather than its stellar photosphere, and modeling it would require a more detailed treatment of the accretion shock than we do here.  The main difference between the four stars in our sample for this paper is, therefore, their range of mass accretion rates (Table \ref{alldiskfits}).

To effectively study the disk and wall structure simultaneously, we need complete spectral energy distributions (SEDs) from optical to millimeter wavelengths. From Paper I, we have quasi-simultaneous, flux-calibrated 0.8--4.5 $\mu$m spectra taken with SpeX on IRTF.  We complement this data with {\it Spitzer} IRS spectra \citep{furlan+06,furlan+11}, optical photometry from \citet{kh95}, near-infrared photometry from 2MASS, IRAC, WISE, and AKARI (accessible through the IPAC Gator service), mid-infrared photometry from MIPS, WISE, AKARI, IRAS, and ISO, and millimeter data from \citet{aw07}, \citet{aw07}, \citet{wendker95}, \citet{guilloteau+11}, and \citet{ricci+10}.

\section{Physical models}

 Here we summarize briefly the main features of the \citet{dalessio+06} disk and dust sublimation wall models, as well as our strategy in testing the wall curvature scenarios described in \S \ref{intro}.

\subsection{Disk structure}
\label{mods}
In the D'Alessio et al. prescription, the temperature and density structures of the disk are calculated self-consistently, assuming heating by stellar irradiation and viscous dissipation.  Viscosity is parametrized through $\alpha$ \citep{shakura_sunyaev73}, which is held constant over the disk.  Accretion is assumed to be steady, and the disk consists of gas and dust.

There are two dust populations in the disk, both of which have grain size distributions with $n(a)=n_0a^{-3.5}$, where $a$ is the grain radius which varies between 0.005$\mu$m and some $a_{max}$.  One of the populations characterizes the disk midplane and has a fixed $a_{max}$ of 1 mm, while the other population describes the upper layers of the disk, with $a_{max}$ allowed to vary.  The populations are vertically distributed as a function of the degree of settling of large grains from the upper layers.   This is parameterized through $\epsilon$, defined as the ratio of the dust-to-gas mass ratio of the upper layers, $\xi$, to $\xi_{standard}$, which is the sum of the mass fractions of the different dust components relative to the gas, i.e. silicates (0.004), graphite (0.0025) \citep{dl84}, and water ice (0.00001).  The value of $\xi_{standard}$=0.00651 is close to $\xi_{ISM}$=0.01 \citep{dl84}.  The dust that settles out of the upper layers of the disk enhances the dust-to-gas mass ratio at the midplane, which is accounted for in this prescription according to Table 3 of \citet{dalessio+06}.  

Opacities for the amorphous graphite, water ice, and silicate dust components with these grain size distributions are constructed from optical constants using Mie theory, assuming that the grains are segregated spheres \citep{pollack+94}.  The optical constants are taken from \citet{dl84} and \citet{warren84}, for the graphite and water ice, respectively, while we used optical constants for several different silicates stoichiometries.   Glassy olivines ($Mg_{2(x)}Fe_{2(1-x)}SiO_4$) and pyroxenes ($Mg_xFe_{1-x}SiO_3$), where $x=Mg/(Mg+Fe)$ indicates the iron content, were taken from  \citet{jaeger+94} and \citet{dorschner+95} and are designated henceforth as D95OlMg(X\%) and D95PyMg(X\%), respectively.  For our crystalline silicates, forsterite and enstatite, we take the best-fitting opacities for those species as determined by \citet{sargent+09}: pure magnesium forsterite ($x$=1) from \citet{chihara+02} and enstatite with $x$=0.9 from \citet{sogawa+06}.  

Input parameters to the code include the stellar properties, the mass accretion rate onto the star, $\alpha$, $\epsilon$, dust composition, $a_{max}$ in the upper layers, the inclination angle $i$, and the outer radius $R_d$.  We have assumed the stellar properties and mass accretion rates derived in Paper I, with the exception of V836 Tau and GO Tau.  For these stars, we found a better match to the combined optical and NIR data with a higher $A_V$ than that found in Paper I (although still within the uncertainty limits), so we correct their stellar parameters to the values given in Table \ref{alldiskfits}.  The inner disk radii were set to the smallest radii of the best-fitting wall model (described below).  The optimal disk structures were then determined by $\chi_{r}^{2}$ fits to fluxes in the whole 0.36$\mu$m to $\sim$3 mm range.


\subsection{Dust sublimation wall}

\subsubsection{Prescription}
\label{dwall}
In the prescription of \citet{dalessio+04} for emission from the inner edge of the dusty disk, dust is assumed to be present in a vertical wall once the disk temperature drops below the dust sublimation temperature, $T_{sub}$, at a radius, $R_{wall}$, for a particular dust composition and grain size distribution.  The dependency of the wall radius on the grain properties is based on the grains' dust absorption efficiency, $q$, given by $q=\kappa^{*+shock}_{P}/\kappa^{wall}_{P}(T_{sub})$.  In this expression, $\kappa^{*+shock}_{P}$ is the Planck mean opacity of the dust at the wavelength range and temperature of the combined stellar and shock emission, and $\kappa^{wall}_{P}(T_{sub})$ is the Planck mean opacity of the dust at the wavelength range and temperature of dust \citep{muzerolle03,mmg02,dalessio+06}.  The relationship between $q$, $T_{sub}$, and $R_{wall}$ can be quantified as \citep[modified from Equation (2) of][]{dalessio+06}:

\begin{equation}
\frac{R_{wall}}{R_{*}}\propto\left(q\right)^{1/2}\left(\frac{T_*}{T_{sub}}\right)^{2}
\label{rwallopa}
\end{equation}

The physical mechanisms leading to curvature in the wall, as described in \S \ref{intro}, influence the assumed values of $T_{sub}$ and $q$.  Micron-size grains will have smaller $q$ than submicron grains, for the same $T_{sub}$ \citep{mmg02}.  Grains of the same size and composition (i.e. $q$) will have higher $T_{sub}$ at higher pressures \citep{pollack+94}.  For the same pressure, and grain size, grains of different composition (e.g. more Fe- or Mg-rich silicates) will have different $q$ values and $T_{sub}$ \citep{gail+99}.  Self-consistent, simultaneous testing of all three wall curvature scenarios would require a more detailed treatment of the wall than we can do here, especially as all three effects are likely to occur at once, to different degrees, and laboratory data on dust sublimation temperatures and opacities under the full range of disk pressures for all likely dust varieties are not always available.  Nonetheless, it is illustrative to use a vertical wall model to test the following questions.  {\it A) Under which conditions {\it can} a vertical wall satisfactorily describe the NIR excess in accreting T Tauri systems?   B) In the cases when it cannot fit the entire excess, which of the three physical mechanisms, if any, most improves the fit?}

To this end, for each T Tauri star we ran a large grid of wall models covering the following range of parameter space for dust sublimation temperatures, sizes, and a small fraction of potential compositions.  We varied $T_{sub}$ from 700 to 1850 K in 50 K increments, $a_{max}$ of the grain-size distribution from 0.1$\mu$m to 20$\mu$m in a non-uniform grid, and silicate compositions consisting of pyroxene and olivine stoichiometries in glassy form with a range of Mg/(Mg+Fe) ratios and Mg-rich crystalline forms, as described in \S \ref{mods}.  Each wall is the \citet{dalessio+05} standard: vertical, with a constant dust grain size distribution, composition, and sublimation temperature.

The total set of models was then compared with the SED of the T Tauri star in two ways.  For our control case, we considered only a single wall and computed the reduced $\chi$-squared value, $\chi^2_r=\chi^2/\nu$, for the entire wall grid.  The number of free parameters, $\nu=N_{SED}-n_{fit}-1$, is large due to $N_{SED}$, the number of wavelengths in the SED, which is approximately 2600.  The number of  fitted parameters, $n_{fit}$, was either 3 or 6 depending on the number of layers in the wall. The wall height was allowed to vary in increments of the gas-pressure scale-height, $H_P$, as $z_{wall}=\xi H_P$ with $0\le \xi \le 4$. In turn, for a given radius $H_P$ is defined as:

\begin{equation}
H_P=R^{3/2}\left( \frac{kT_c}{GM_*\mu(T_c,\rho_c)m_H} \right)^{1/2}
\label{scaleheight}
\end{equation}

\begin{figure*}
\includegraphics[angle=0, scale=0.9]{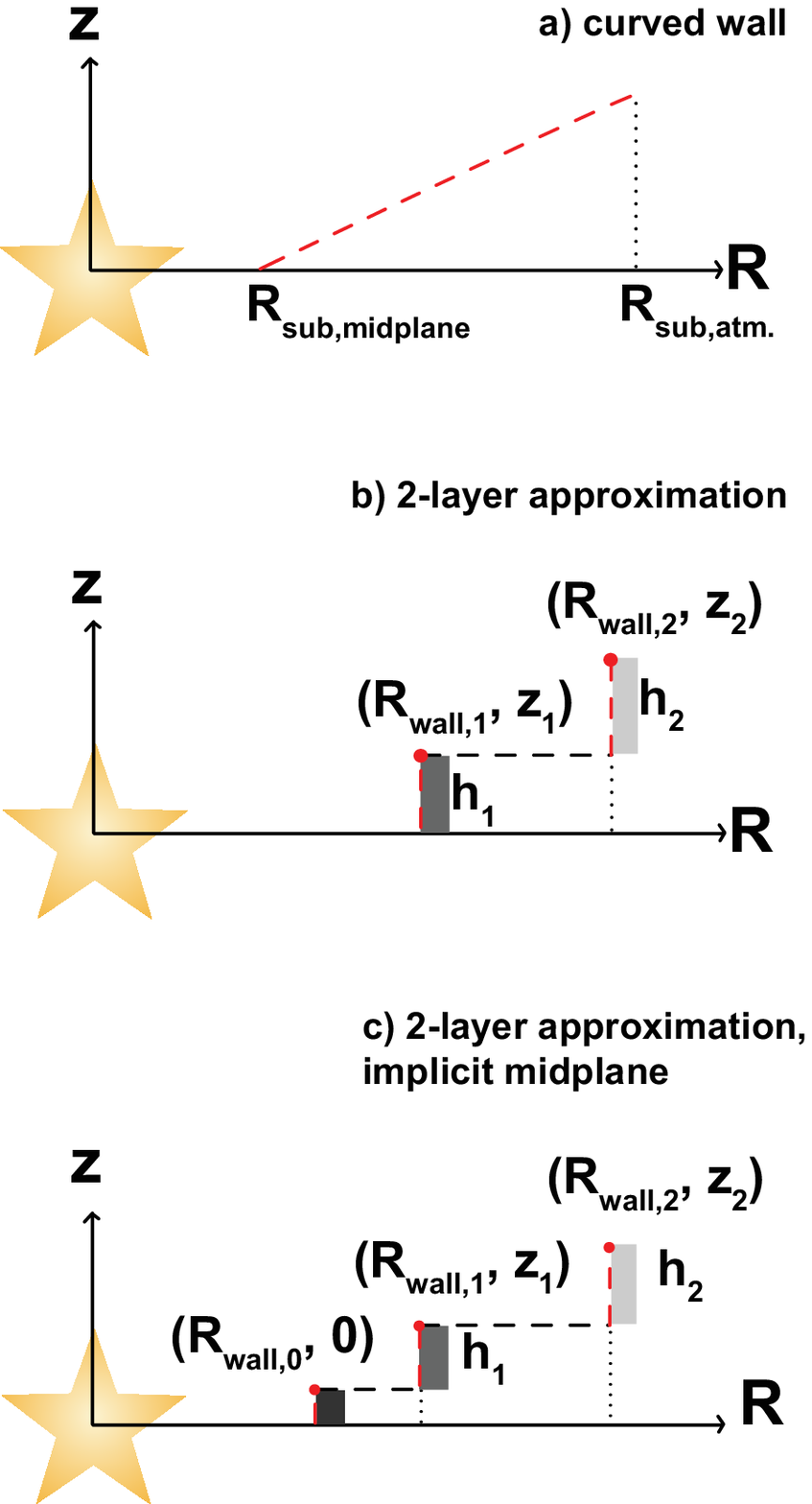}
\caption{Wall curvature: \emph{a)} Simple curved wall, \emph{b)} A first-order approximation of the curved, with two vertical layers (to contain two dust populations), \emph{c)} It may be that the two layers we detect do not probe down to the midplane, in which case there should be a third, thin layer at the midplane.  \label{twolayer}}
\end{figure*}

\noindent The other case tests the wall curvature mechanisms through the use of a two-layer wall, in which each layer is vertical but there is a radial offset, like a step function, as shown in Fig. \ref{twolayer}.  We allow the disk component to extend in to the radius of the lower wall layer, since the contribution of the disk between the two wall layers is small ($\sim$1/10) compared with the frontally illuminated wall layers.  We assumed the best-fitting single-wall dust composition for the bottom layer, allowing $T_{sub}$, $a_{max}$, and $\xi$ to vary in the bottom layer and the dust composition, $T_{sub}$, $a_{max}$, and $\xi$ to vary in the top layer, with the requirement that the $\xi_{upper}$ represents the height $z_{2}^{'}$ in Fig. \ref{twolayer}.   The resulting best-fits are given in Table \ref{alldiskfits}.  

Briefly, while our two-layer wall is obviously intended as an approximation it is worth noting several physical effects that may cause second-order curvature.  First, in reality there should be smooth distribution of maximum grain sizes, rather than bimodal populations.  Additionally, the reallocation of the sublimated dust into the gas phase of the upper layers would reduce the disk's opacity in the `flat' region between the two wall layers, cooling it below the value in our models.  However, the gas in the upper layers close to the star should also be heated via direct irradiation by high-energy photons.  The combination of these effects likely results in a $\tau$=1 surface between the wall layers that is not flat, decreasing the amount of direct illumination of the upper layer.  However, any such curvature would result more direct illumination of the (formerly) flat region, ultimately compensating for the decreased flux of the upper wall layer, especially over a wider variety of inclination angles \citep{isellanatta05}.

\subsubsection{Detailed example of CI Tau}
\begin{figure*}
\includegraphics[angle=0, scale=0.9]{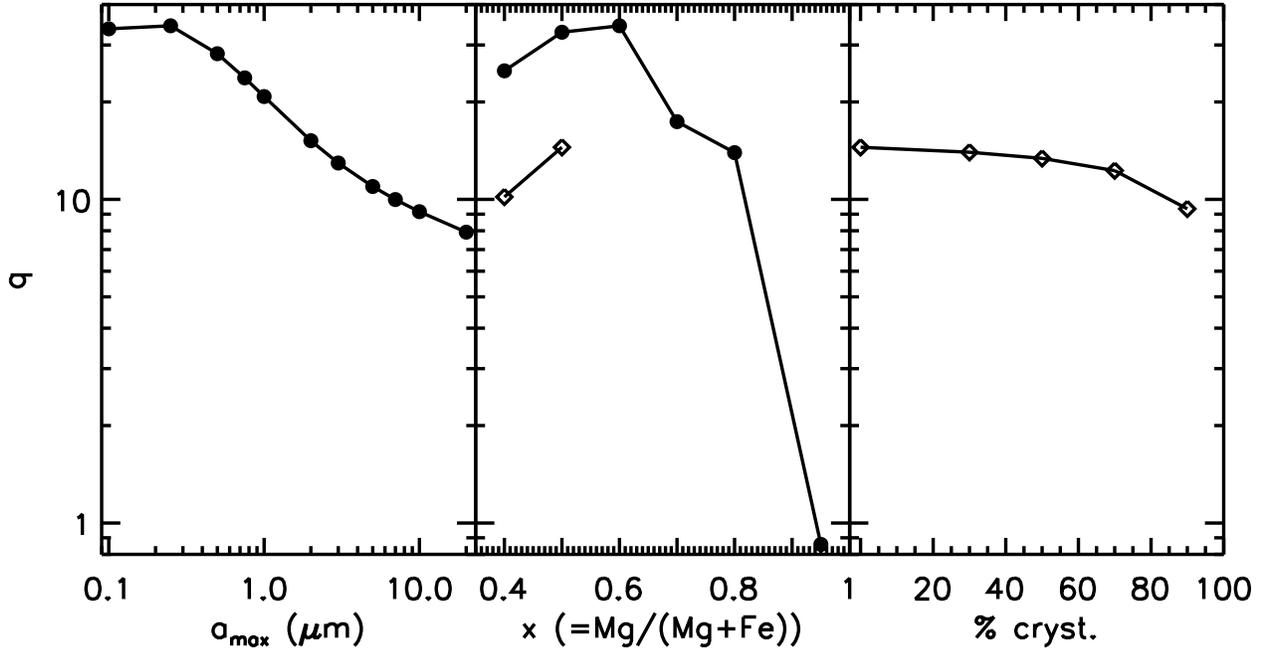}
\caption{Comparison of the variation in the dust absorption efficiency, $q=\kappa^{*+shock}_{P}/\kappa^{disk}_{P}(T_{sub})$, for the wall dust population when individual dust properties are varied.  In all cases, the dust sublimation temperature is held constant at 1600 K. {\it Left:}  Change in $q$ as a function of $a_{max}$, for iron-rich pyroxene dust (D95PyMg60).  {\it Middle:} Change in $q$ as a function of iron content for pyroxenes (D95PyMgX, filled circles) and olivines (D95OlMgX, open diamonds), for $a_{max}$=0.25$\mu$m.  {\it Right:} Change in $q$ as a function of crystallinity for a mixture of iron-rich olivine (D95OlMg50) and pure forsterite, for $a_{max}$=0.25$\mu$m. We note that in this case, changing the crystallinity results in a de facto change in the iron content, as we are mixing an iron-rich amorphous olivine with an iron-free crystalline olivine. \label{oparatios}}
\end{figure*}

To demonstrate better our analysis, we show in depth the process of fitting the wall of CI Tau, the star with the largest excess. First, we demonstrate how the shape and absolute flux of the wall emission varies with different parameters.  In Fig. \ref{oparatios} we demonstrate how $q$ varies as a function of the grain size distribution, silicate stoichiometry, silicate iron-content, and silicate crystallinity.  We recover the expected behavior that small grains are more efficient absorbers than large grains over our wavelength ranges.  Additionally, olivine is a factor of $\sim$2.5 lower in $q$ than pyroxene of a comparable iron content and $a_{max}$.  Although the true absorption of olivine is greater than that of pyroxene at all three wavelength regimes (shock, star, disk), pyroxene is more efficient at retaining what it absorbs.  Also note that pyroxene absorbs most efficiently with an iron content of 40 to 50\% relative to Mg ($x=0.5-0.6$).  The models with larger $q$ will have larger radii, by Eq. (\ref{rwallopa}).  Because of the dependence of $H_P$ in Eq. (\ref{scaleheight}) on $R$, for the same sublimation temperature, these models will also produce a larger value of $H_P$.  Therefore it will require a smaller scale factor to reproduce a given solid angle than will grayer dust.

\begin{figure*}
\includegraphics[angle=0, scale=0.9]{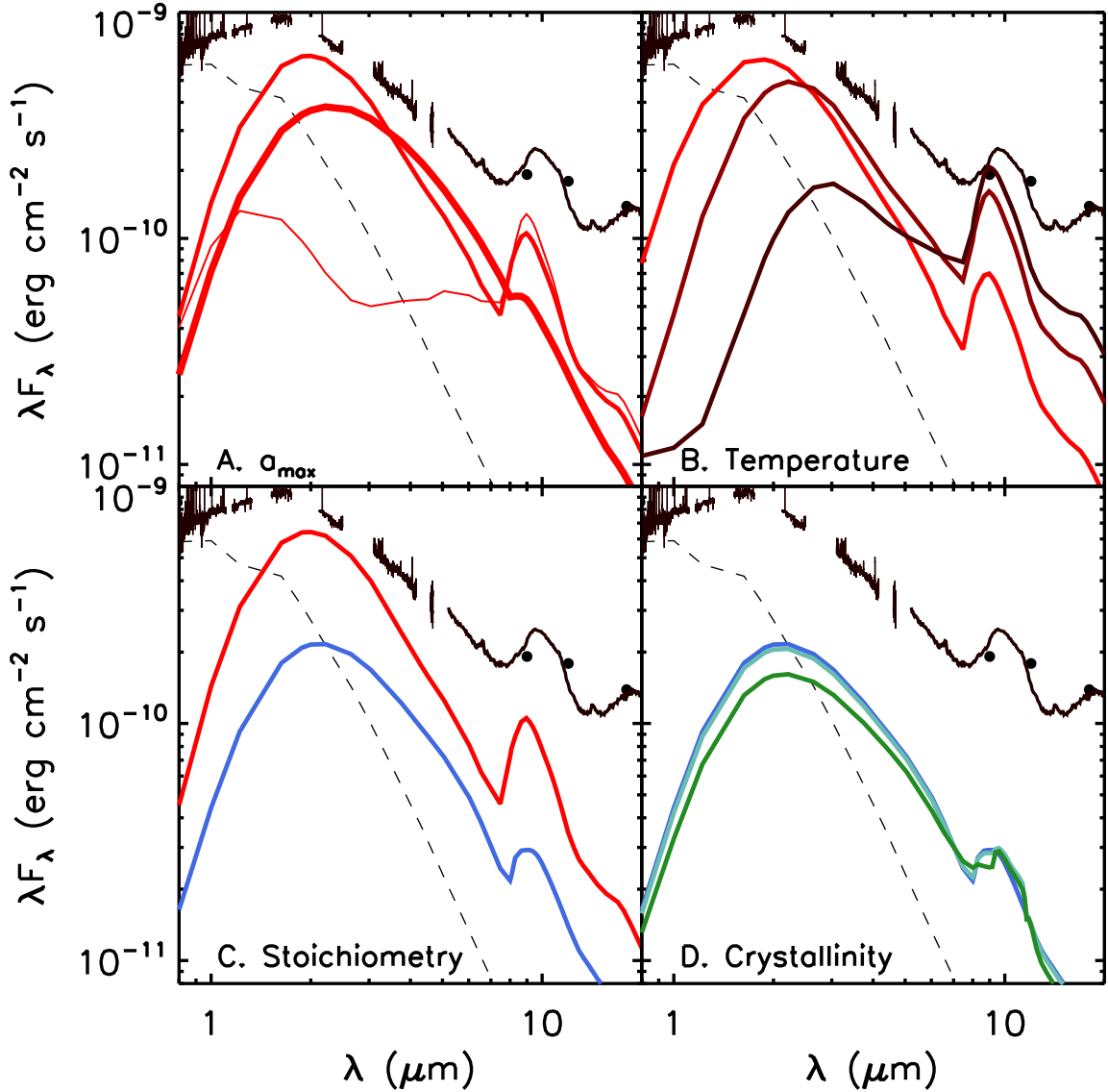}
\caption{Comparison of wall emission for CI Tau for models in which the following parameters are varied. {\it A. Grain-size:} Pyroxene models (D95PyMg50) at 1600 K with $a_{max}$ of 0.25 (thin, red line), 1.0 (red line), and 10 $\mu$m (thick, red line).  {\it B. Temperature:} Pyroxene models (D95PyMg50) of $a_{max}$ 1.0 $\mu$m at sublimation temperatures of 1800 (red line), 1300 (darker red line), and 900 (darkest red line) K. {\it C. Stoichiometry:} Models with $a_{max}$ of 1.0 $\mu$m at 1600 K of pyroxene (D95PyMg50, red line) and olivine (D95OlMg50). {\it D. Crystallinity:} Olivine (D95OlMg50) models with $a_{max}$ of 1.0 $\mu$m and increasing fractions of forsterite: 0\% (blue), 50\% (blue-green), and 90\% (green).  See caption to Fig. \ref{oparatios} for caveat on the iron content of panel D.  All models are 4$H_P$ in height. \label{wall_pyr_oli}}
\end{figure*}

In addition to producing variation in the radius and emitting area, changing the dust properties affects the shape of the wall emission, as demonstrated in Fig. \ref{wall_pyr_oli} for a set of models in which the wall height is fixed at 4$H_P$ where $H_P$ varies from model to model with the values of $R_{wall}$ and $T_{sub}$.   For models of different grain sizes, the effect on the emission is most pronounced between 0.25 and 1 $\mu$m.  Since the wall has an atmosphere, the optically thin emission comes from temperatures between $T_{sub}$ and the temperature corresponding to the $\tau$=2/3 depth, $T_{eff}$.  

For large $q$ values, i.e. submicron grains, most of the radiated energy is absorbed close to the surface and $T_{eff}$ is substantially less than $T_{sub}$, leading to relatively flat NIR emission with two broad bumps at the wavelengths corresponding to $T_{sub}$ and $T_{eff}$, i.e. 1 to 2 $\mu$m and 3 to 7 $\mu$m, respectively.  Emission is seen at 10$\mu$m because the silicates have a high optical depth there than in the continuum.  As the value of $a_{max}$ increases to micron scales, $T_{eff}$ approaches $T_{sub}$, leading to more blackbody-shaped emission, which shifts slightly to longer wavelengths with larger $a_{max}$.  Variations in $T_{sub}$ can mimic this observed wavelength shift in peak emission, depending on the grain size.  There is also a noticeable difference in the peak flux and shape of the 10 $\mu$m feature between glassy olivine and pyroxene with the same Mg/(Mg+Fe) ratio.  Because the sublimation radius of pyroxene is farther from the star than that of olivine, its flux is a factor of 3 higher over the NIR than the olivine model.  In contrast, increasing the crystallinity of the silicate grains affects mainly the shape of the 10 $\mu$m silicate complex.  We note that for pyroxene, changing the iron content affected the absolute flux more than the shape of the emission, so we did not include a figure with that comparison.

\begin{figure*}
\includegraphics[angle=0, scale=0.7]{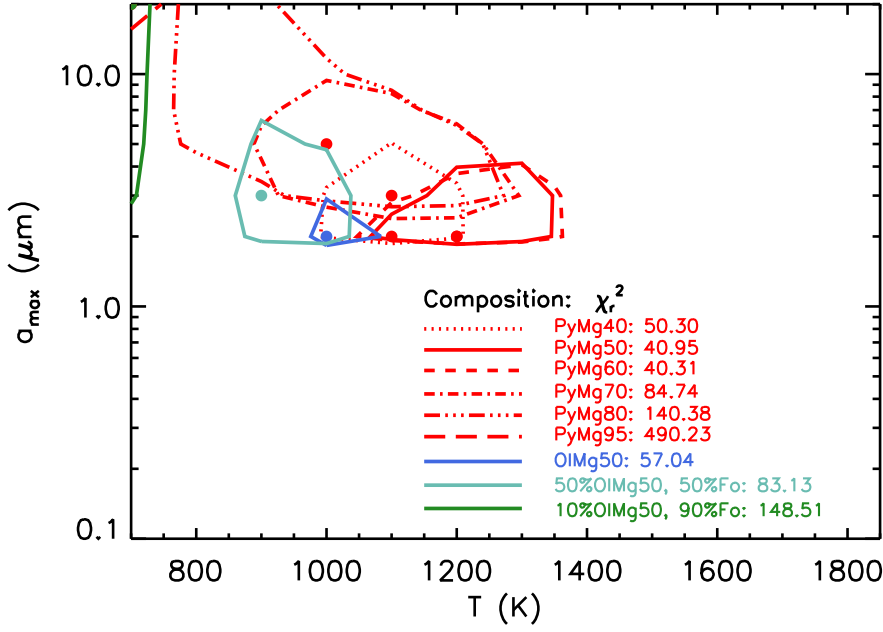}
\caption{Reduced $\chi^2$ for the vertical wall models of CI Tau as a function of $a_{max}$ and $T_{sub}$, with the 5 to 40 $\mu$m region weighted by 0.1.  The range of $a_{max}$ is 0.1, 0.25, 0.5, 0.75, 1.0, 2.0, 3.0, 5.0, 7.0, 10.0, and 20.0 $\mu$m, while the range of $T_{eff}$ is 700 to 1800 K in increments of 50 K.  The last free parameter was the scale factor, $\xi$ = $z_{wall}/H_P$, which was varied from 0 to 4.  Compositions are labeled in the legend, along with the minimum $\chi_{r}^{2}$ for that composition. The best-fitting $a_{max}$ and $T_{eff}$ are indicated for each composition by a solid circle.  Lines are 3$\sigma$ confidence intervals.  The overall best-fit is for pyroxene with $x=Mg/(Mg+Fe)$ of 0.6, $a_{max}$ of 2 $\mu$m, and $T_{eff}$ of 1200 K (short-dashed, red line).  \label{chi_1wall}}
\end{figure*}

To test our control case, the single vertical wall, we added the wall emission to that of the stellar photosphere, assuming the colors of \citet{kh95}, and the best-fitting disk model for wavelengths $>$20$\mu$m. We assume a physically motivated upper limit to the wall scale factor, $\xi=z_{wall}/H_P$, of 4 since this is typically the height at which most of the stellar radiation would be absorbed when it is incident at an angle, and allow $\xi$ to vary for each model from 0 up to this upper limit in increments of 0.1.  Then we computed the reduced $\chi^2$, $\chi_{r}^{2}$ to the SED over the entire wavelength range, weighting the IRS spectrum by a factor of 0.1 to account for the lower spectral resolution compared with SpeX.  The best-fitting $T_{sub}$ and $a_{max}$ are plotted in Fig. \ref{chi_1wall}, and we note several significant points.  First, within 3$\sigma$, none of the best-fits for any of the compositions and any of the temperatures came from grains with $a_{max}$ less than 2 $\mu$m.  Second, the best-fitting iron content was between 60 and 40\%.  Finally, the best-fitting temperature was between 900 K and 1200 K.

The variation in the $\chi_{r}^{2}$ fit between compositions is driven by two factors: the shape and absolute flux of the emission excess between 1.5 and 4 $\mu$m and the shape of the short wavelength side of the IRS spectrum.  Within the 4$H$ upper limit to the wall height, the only material that can reproduce the absolute 1.5 to 2$\mu$m {\it and} the 2 to 4$\mu$m excess is large-grained pyroxene.  A wall with olivine grains cannot reproduce the absolute flux of the excess without a wall height of at least 5$H$, and its shape is too flat over that region, as demonstrated in Figure \ref{wall_pyr_oli}, even at 1600 K.  Additionally, the larger pyroxene grains nicely match the shape of the 8 to 9$\mu$m side of the 10 micron feature, although the absolute flux there is too low.  Overall, the composition with the smallest $\chi_{r}^{2}$ ($\sim$90) was 2$\mu$m, 100\% amorphous pyroxene with 40\% iron content at 1200K.

\begin{figure*}
\includegraphics[angle=0, scale=0.7]{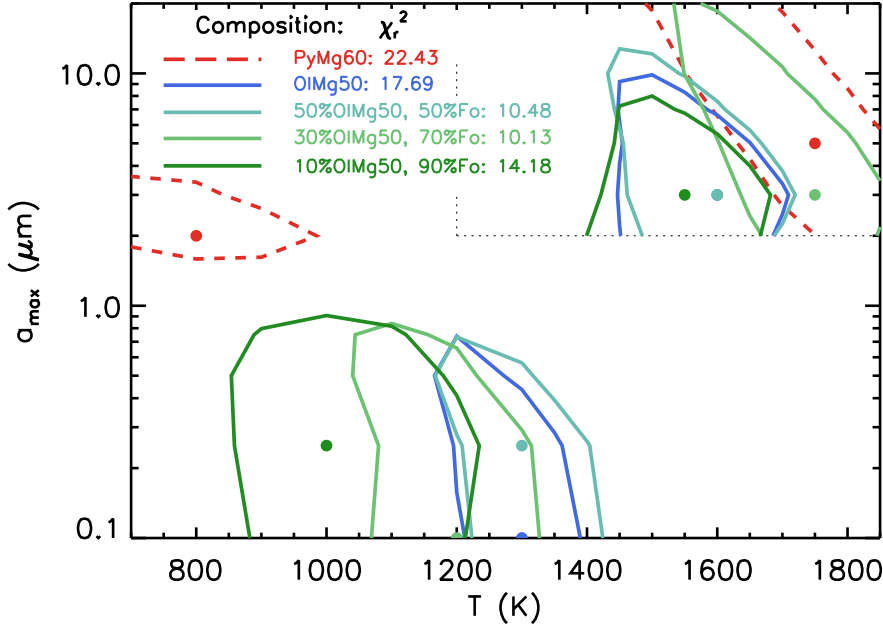}
\caption{Analogous plot to Fig. \ref{chi_1wall}, but for the two-layer wall models.  The lower layer of the wall has its composition fixed as pyroxene with $x=0.6$, and its grain size and temperature are limited to the parameter space greater than or equal to the best-fitting single wall model (i.e. the area enclosed by the black, dotted lines). Listed compositions are for the upper layer of the wall. The overall best-fit for the upper layer is for amorphous olivine with $x$=0.5, $a_{max}$ of 0.25 $\mu$m, and $T_{eff}$ of 1200 K.  \label{chi_2wall}}
\end{figure*}

For the two-layer wall case, we assumed that the bottom layer was comprised of the best-fitting single wall composition with $a_{max}>$2$\mu$m and $T_{sub}>$1200K, and allowed the top layer to vary over the entire range of compositions, temperatures, and $\xi$.   The composition of the upper layer primarily affected the shape of the 10 $\mu$m feature and the shape of the continuum from 5 to 7 $\mu$m.  Models with pyroxene underfit the 5 to 7 $\mu$m region as well as the 9.4 to 10 $\mu$m emission, since pyroxene peaks at 9.3 $\mu$m.  Taking the upper layer to contain submicron olivine solves those problems.  However, regardless of the grain size, adding only amorphous olivine still underfits the longer wavelength side of the 10$\mu$m feature; forsterite fractions between $\sim$ 50 to 70\% are required to match that part of the IRS spectrum.  Separate from the grain-size or composition, the upper layer must also have a lower $T_{sub}$ than the lower layer.  The resulting best-fit $\chi_{r}^{2}$ are displayed in Fig. \ref{chi_2wall}.  It is clear that the two-layer approximation to a curved wall fits the SED far better than a vertical wall; the worst-fitting two-layer wall model $\chi_{r}^{2}$ is almost a factor of two smaller  than the best-fitting vertical wall model, even though they have the same grain composition.


\section{Results}

\subsection{Evidence of wall curvature}
\label{wallfits}

\begin{figure}
\includegraphics[angle=0, scale=0.7]{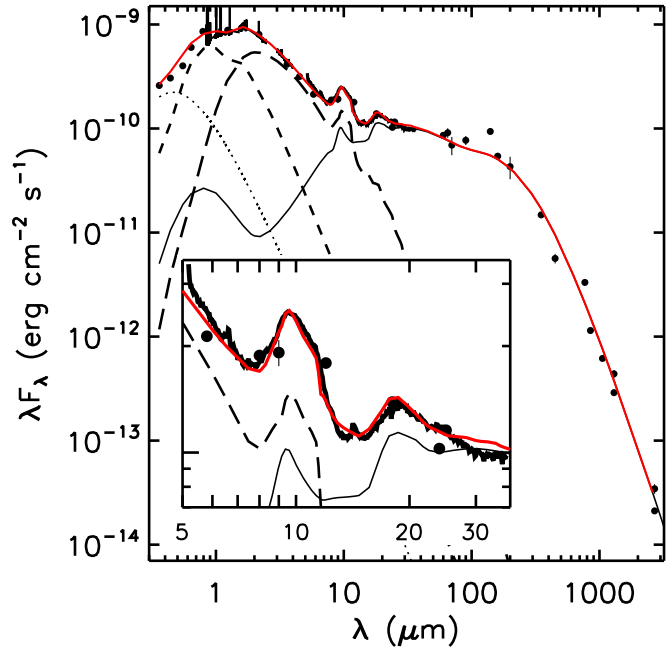}
\caption{Model fit to CI Tau SED.  Solid black lines are the SpeX and Spitzer IRS data.  Photometry is indicated with solid black circles; error bars are also plotted but generally are smaller than the plot symbols.  Thick, dashed, black line is the photosphere used by the model.  Dotted black line is a simple blackbody at $T$=8000K, added to fit roughly the optical excess produced by the accretion shocks.  Long-dashed line is the 2-layer dust sublimation wall.  Thin solid black line is the disk (note the scattering and thermal components).  The composite model is the solid red line. The inset shows an enlargement of the 10 $\mu$m region.  Photometry taken from AKARI IRC\citep{ita+10}, AKARI FIS, the IRAS SSC, the ISO archive, \citet{aw05}, \citet{wendker95}, and \citet{guilloteau+11}. \label{citau_sedfit}}
\end{figure}

\begin{figure}
\includegraphics[angle=0, scale=0.7]{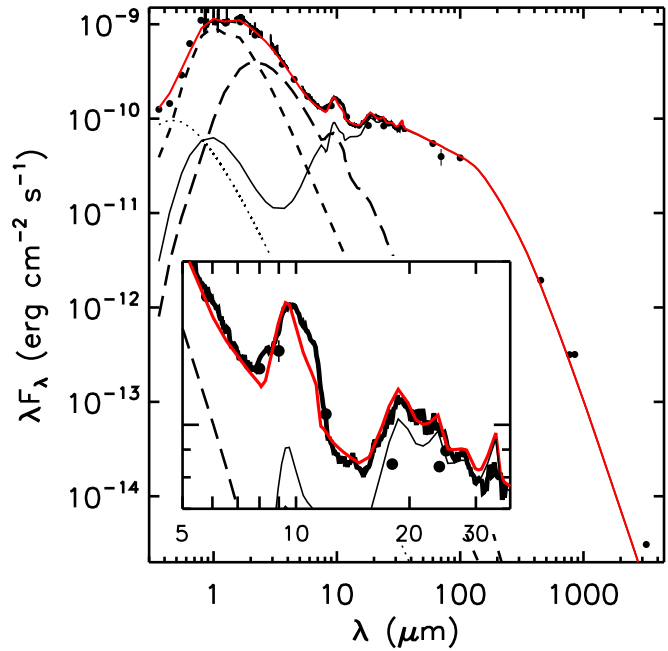}
\caption{Model fit to DE Tau SED. Components are labeled as in Fig. \ref{citau_sedfit}. Photometry taken from the {\it Spitzer} Legacy Science Program Taurus Catalog, AKARI IRC \citep{ita+10}, the IRAS FSC, \citet{aw05}, \citet{wendker95}, and \citet{ricci+10}. \label{detau_sedfit}}
\end{figure}

\begin{figure}
\includegraphics[angle=0, scale=0.7]{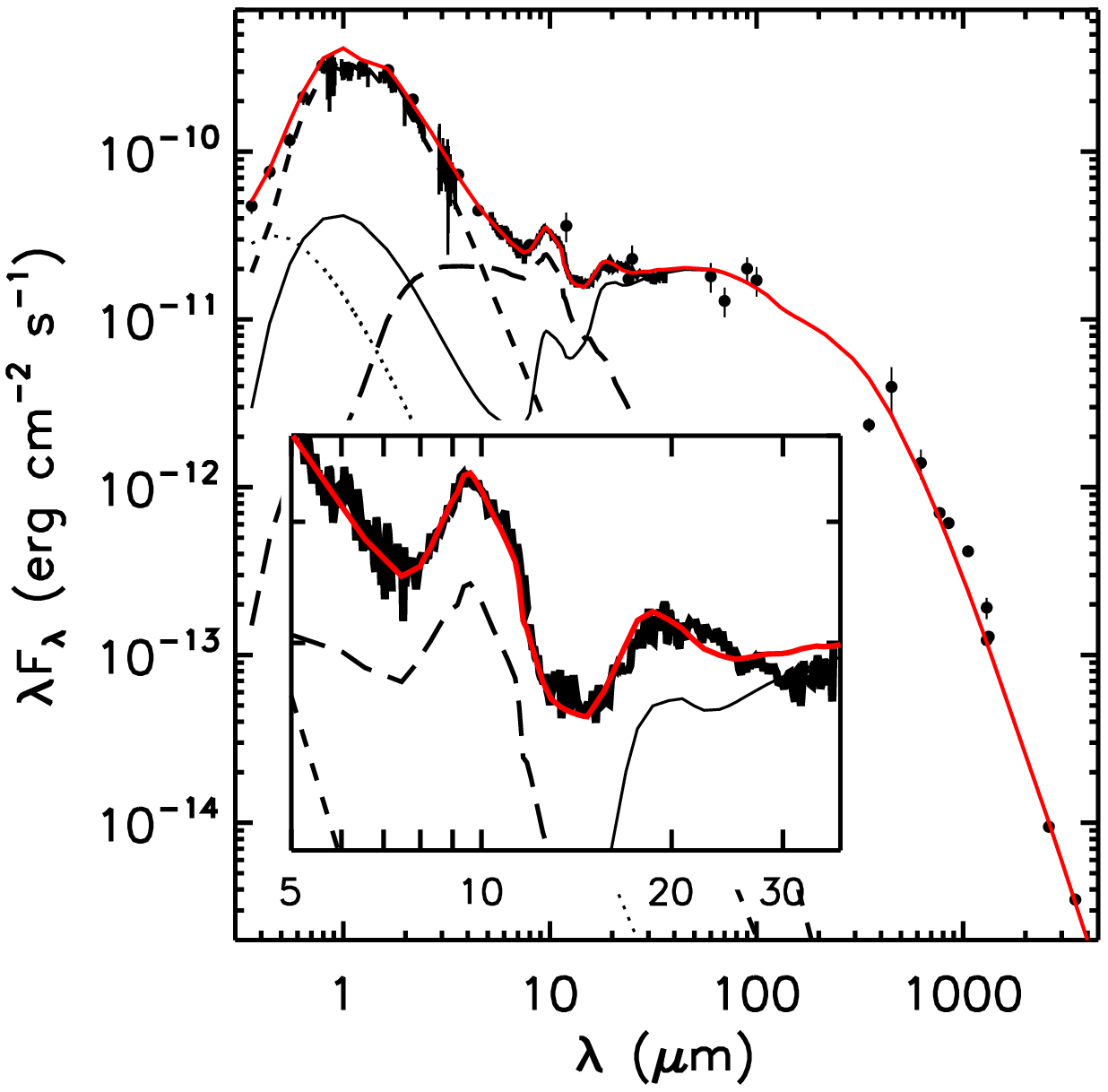}
\caption{Model fit to GO Tau SED.  Components are labeled as in Fig. \ref{citau_sedfit}. Photometry taken from AKARI IRC\citep{ita+10}, AKARI FIS, the IRAS SSC, the ISO archive, \citet{aw05}, \citet{wendker95}, and \citet{guilloteau+11}. \label{gotau_sedfit}}
\end{figure}

\begin{figure}
\includegraphics[angle=0, scale=0.7]{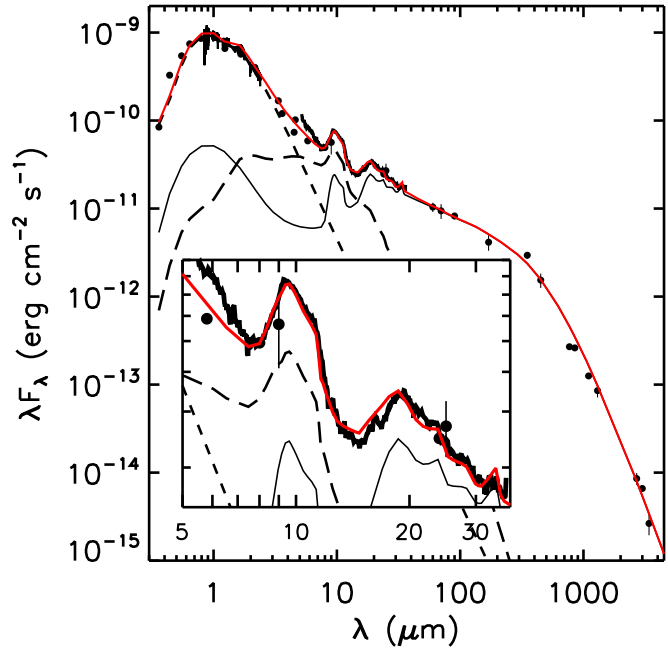}
\caption{Model fit to V836 Tau SED.  Components are labeled as in Fig. \ref{citau_sedfit}. Photometry taken from AKARI IRC\citep{ita+10}, AKARI FIS, the IRAS SSC, the ISO archive, \citet{aw05}, \citet{wendker95}, and \citet{guilloteau+11}. We note that the variability between the IRAC photometry and Spitzer IRS spectrum from 5 to 7$\mu$m is likely real and we chose to fit the photometry. \label{v836tau_sedfit}}
\end{figure}

The best-fitting models are shown in Figs. \ref{citau_sedfit} to \ref{v836tau_sedfit}.  All of the disks in the sample were better fit by a two layer wall than a single layer wall, and it appears that all three physical mechanisms (grain size, pressure dependence, and grain composition) play a role.  For each of the disks, $\sim$1000 K submicron grains in the upper layers were required to fit the 10$\mu$m feature of the IRS spectrum and the surrounding continuum emission.  A lower layer with a combination of larger, hotter grains was able to fit the 2 to 5 $\mu$m continuum emission, while grains the size of those in the upper layer were not, even at a higher temperature.  This result is consistent with with previous wall studies by \citet{isellanatta05} and \citet{ajay+07} in which the effects of pressure and grain size segregation were taken into account.  

However, we note that our narrow selection criteria may reveal a connection between the accretion properties and inner disk properties.  The lower layer grain size and temperature appear to decrease as a function of mass accretion rate.   For CI Tau and DE Tau, with \Mdot $\sim$ 2$\times$10$^{-8}$\Msun/yr, the lower layer of the wall has $T_{sub}\sim$1700K and $a_{max}\sim$3$\mu$m.  GO Tau and V836 Tau, with \Mdot$\leq$4$\times$10$^{-9}$\Msun/yr, have lower layers with $T_{sub}\sim$1250K and $a_{max}$ between 0.5 and 1$\mu$m.  In addition, while the $T_{sub}$ ranges for each wall layer are roughly consistent with sublimation temperatures for silicates at the pressures indicated by our disk model for each star, there are interesting implications for the high temperature of the lower layer, which we discuss in \S \ref{wallsettle}.  

We also find evidence in support of the third curvature mechanism: variation in the composition of the grains, in this case the silicates.  The only lower layer walls that were able to reproduce the 2 to 5 $\mu$m excess in CI Tau and DE Tau with a height of at most 4$H_P$ had a pyroxene stoichiometry with $x\sim$0.5$-$0.7.  Olivine walls of a similar iron content required heights of at least 6$H_P$ in order to match the absolute flux at 2$\mu$m.  Likewise, pyroxene in the upper layer produced a poor fit to the 10$\mu$m silicate feature, as the wavelength at which that feature peaks was shifted to $\sim$9.3$\mu$m in the models while the the feature in the data peaked closer to 9.8$\mu$m.  The best fits for the upper layers were achieved for a 100\% olivine composition divided into $\sim$40\% to 30\% amorphous grains and $\sim$60\% to 70\% Mg-rich forsterite.  It is less easy to distinguish between a pyroxene or olivine stoichiometry for the lower wall of the weaker accretors, GO Tau and V836 Tau, as their best-fitting scale factors, $\xi$, are 1$H_P$ or less, and in their case the shape of the 10$\mu$m feature is almost solely determined by the upper layer grain properties.

\subsection{Settled outer disk structure}
\label{diskfits}

The values of $\alpha$ and $\epsilon$ were constrained primarily by the quality of the fits to the submillimeter data, while the dust properties affected mainly the fit to the Spitzer IRS spectra.  All of these TTS disks were best-fit with $\epsilon \leq$ 0.01, consistent with the comparison of spectral indices with a D'Alessio et al. model grid by \citet{furlan+06} and indicating substantial dust settling.  However, there is no correlation between the degree of dust settling and the mass accretion rate.  Additionally, three of the four disks have $\alpha <$ 0.01, the canonical value.  This may be a selection effect, as we required the disks to have millimeter data, and the disks are average-to-massive at 2$\times$10$^{-3}$ $-$ 7$\times$10$^{-2}$ \Msun \citep{aw05}.

For each of the stars in our sample, the disk contribution to the 10 $\mu$m feature is at most 50\%. While the upper wall layers have uniformly submicron grains, the upper disk layers are well fit by dust populations with a range of grain sizes from 0.25$\mu$m to 3$\mu$m.  In general the 40 to 100$\mu$m regions are best-fit by submicron grains, but two of the disks have 20 micron features that are well-fit by micron-sized grains.  The crystallinity in the disk is also less than in the wall; the regions of the IRS spectra beyond 20 $\mu$m are well-fit by a spatially uniform crystalline fraction of 10\%.  However, in reality there is likely a radial gradient between the upper layer of the wall and the disk in terms of the crystalline fraction.  If we increased the crystallinity in the whole disk to match the 10 $\mu$m features, we then overproduced the crystalline features from 20 to 35 $\mu$m.  Pyroxene grains provided the best match with the 20$\mu$m silicate feature.  Regarding the iron content of the disk silicates, unlike the wall models we note that disk models with different Mg/(Mg+Fe) ratios do not differ enough for us to be sensitive to variation in $x$ given the uncertainty in the IRS spectra and dust opacities.  We assumed a value of $x=$0.8, consistent with the average $x$ found in solar system bodies originating between 1 and 5 AU \citep{nakamura+11,zolensky+06}.  The slope of the submillimeter data is better fit by a dust population with $a_{max}$=1mm than by any smaller size.  DE Tau shows signs of additional emission around 3mm that could reflect a change in the midplane dust population either globally or as a function of radius or free-free emission \citep{loinard+07}.  There are no measurements of this system at cm wavelengths, so we are unable to address free-free emission.  Since we are concerned mainly with the inner disk here, we do not seek to improve the fit to this single point.  

\section{Discussion}

\subsection{Shadowed disks vs. settled disks}

Comparing our curved wall approximations with the physical disk structure models can inform one of the current questions surrounding the role of the dust sublimation wall on the outer disk structure, whether the wall is `puffed-up' and shadows the outer disk \citep{natta+01,dullemond+01} or if grains in the wall and disk have settled down to the midplane, producing less flaring of the whole disk \citep{dd04b,dalessio+06}.  Shadowing of the disk by the wall has been used to explain the classification of Herbig Ae stars into two groups based on the slope of their SEDs; flat SEDs are considered to be flared, and SEDs with a more negative slope are considered to be self-shadowed by their walls \citep{meeus+01,dd04}.  The models applied to these SEDs are, however, passive models that do not account for the effects of accretion on the surface density of the disk.  In contrast, the D'Alessio et al. models are self-consistent, irradiated accretion disks.  This is particularly important in the inner part of the disk directly behind the wall, as the midplane of accreting stars is typically at the dust sublimation temperature due to viscous heating from the inner edge out to several tenths of AU \citep{dalessio+99,dalessio+01,dalessio+06}, so $H_P$ in Equation (\ref{scaleheight}) is the same behind the wall as in the wall itself \citep{cd11}.

\begin{figure}
\includegraphics[angle=0, scale=0.7]{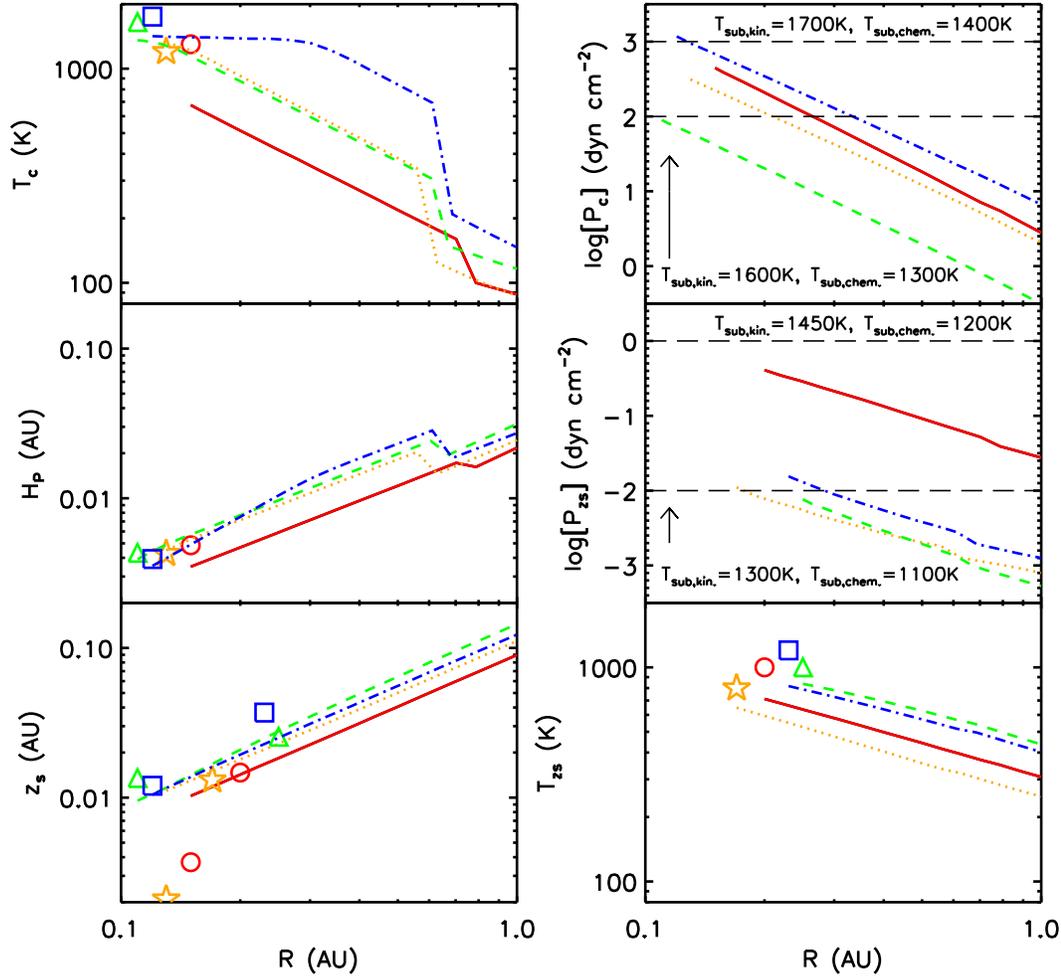}
\caption{Comparison of inner disk structure with wall:  \emph{Left-top)} Midplane temperature ($T_c$), \emph{Left-middle)} gas-pressure scale height ($H_P$), \emph{Left-bottom)} disk surface ($z_s$), \emph{Right top)} midplane pressure ($P_c$), \emph{Right middle)} surface pressure ($P_{zs}$), and \emph{Right-bottom)} surface temperature ($T_{zs}$) for the inner disks of CI Tau (blue dot-dashed line, square symbol), DE Tau (green dashed line, triangle), GO Tau (orange dotted line, star), and V836 Tau (red solid line, circle). Symbols are plotted at with the temperatures and radii of the lower and upper wall layers.  In the two pressure panels, the black labels state the sublimation temperatures for olivine dust in either kinetic or chemical equilibrium at the pressures indicated by the long-dashed, black lines.  \label{diskstructs}}
\end{figure}

Diagnostic plots of the temperatures and pressures in the midplane and disk surface, as well as the gas pressure scale height and disk surface as a function of radius are given in Fig. \ref{diskstructs}.  In particular, the left hand side of Fig. \ref{diskstructs} shows the midplane temperatures ($T_c$), gas-pressure scale height ($H_P$), and the disk surface where $\tau=1$ to the stellar radiation ($z_s$) as a function of disk radius for our sample.  For all the disks, $T_c$ decreases steadily out to $\sim$0.7 AU and then drops off sharply before leveling out as heating via stellar and shock irradiation becomes dominant over viscous heating.  $H_P$ rises monotonically from the wall outward in radius, except for a small dip near $\sim$0.7 AU where the snowline intersects the midplane.  The wall heights are also shown overplotted on the disk surface, $z_s$, and compared with $z_s$ in Table 1.   It is clear that the wall heights are consistent with the $z_s$ for each disk, within a factor of 1.5, and for the cases in which the wall is slightly higher than $z_s$ of the disk, it could shadow at most the region of the disk within 0.1 AU immediately behind the wall, not enough to affect the structure of the bulk of the disk. Since the surface height is where most of the stellar radiation is absorbed at any given radius, if the wall is not higher than this surface, it cannot shadow the disk behind it. The effects commonly attributed to shadowing are equally well described by settling, as demonstrated by our fits to the disks in this sample, which span two orders of magnitude in our settling parameter.

\subsection{Grain fragmentation, settling, and dust-gas ratio enhancement in the wall}
\label{wallsettle}

The wall presents a cross section of the disk, allowing us to observe the dust populations in various layers, including the only direct view of the midplane at infrared wavelengths.  Our results indicate settling of the dust in the inner disk; the absence of large grains in the upper layer of the wall combined with their presence in the lower layer for the high accreters suggests that the larger grains were removed from the upper layers by settling. However, the grain sizes in the lower layers of the lower accreters are not significantly smaller than the grain sizes in their upper layers, in contrast with the pattern seen for the higher accreters.  There are at least two physical mechanisms that can explain this result.  First, we could suppose that there is a third, much thinner layer in the wall at the location of the true midplane where the largest grains are concentrated.  If the grains are big enough, this layer would lie closer to the star and have a much smaller emitting area relative to the upper two layers (see bottom panel of Fig. \ref{twolayer}), making it impossible to separate from the emission of these upper layers in the NIR excess using our simple approximation to a curved wall.  If our higher mass accretion rate disks are more turbulent, then even for a settled wall the micron-sized grains from the midplane wall layer could be lifted into the next highest wall layer, producing the result we see.  If the lower mass accretion rate disks are less turbulent, then the settled dust grains could remain in the midplane layer, leaving similarly sized submicron grains in the other two layers.

Alternatively, our results could indicate that we are seeing fragmentation limited grain growth.  According to \citet{birnstiel+12}, in fragmentation limited regions of the disk, there is a maximum size to which grains can grow before being halted by erosion via turbulent fragmentation, as given by their Eq. 8: 
\begin{equation}
a_{frag}=f_f\frac{2}{3\pi\alpha_t}\frac{\Sigma_g}{\rho_s}\frac{u_f^2}{c_s^2}
\label{birnstiel8}
\end{equation}
\noindent 

In this equation, $\alpha_t$ is the turbulent $\alpha$-parameter, $\Sigma_g$ is the gas surface density, $\rho_s$ is the internal dust grain density, $u_f$ is the fragmentation threshold velocity, and $c_s$ is the sound speed.  The maximum grain size depends on the disk temperature structure through the sound speed, with hotter disks producing a smaller $a_{frag}$. If this $a_{frag}$ is small enough, say on the order of a few microns, its depletion height would still be high enough in the disk that we could see them in our NIR excess.  By extension, when the disk temperature decreases to the point that fragmentation no longer prevents grains from growing larger than a few microns, these newly formed bigger grains should have a lower depletion height, putting them in the less detectable midplane layer and leaving only submicron grains in the upper layers of the wall.

\begin{figure}
\includegraphics[angle=0, scale=0.7]{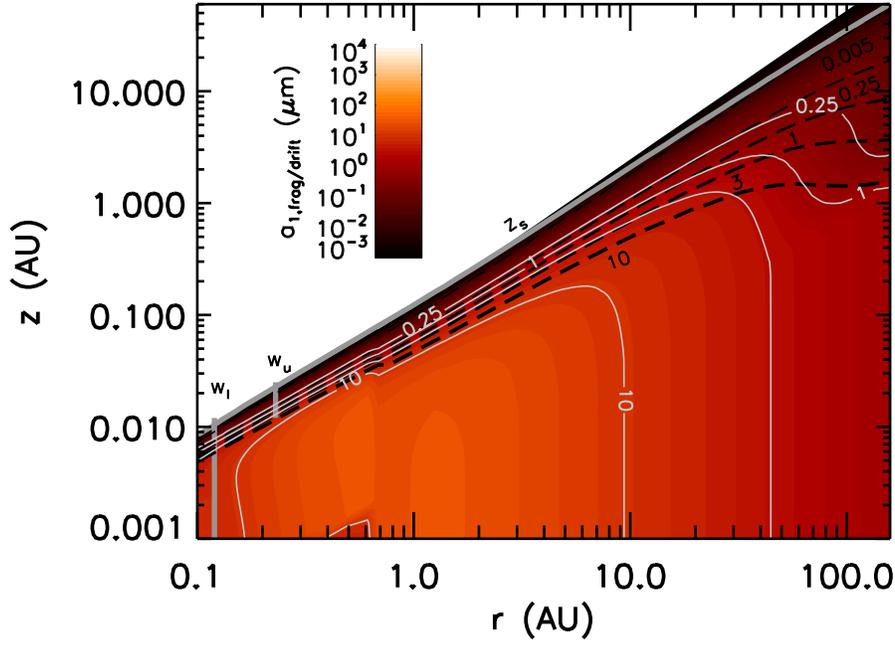}
\caption{CI Tau: Maximum grain sizes predicted by Eq. \ref{birnstiel8}, given our model temperature and density structures and $\alpha$ as input.  Solid white contours are selected maximum grain sizes in microns.  Dashed black lines are predicted depletion heights for the same set of grain sizes from settling theory \citet{dd04b}. The two wall layers and the disk photosphere are indicated by thick, gray, labelled lines.  \label{citau_frag}}
\end{figure}

\begin{figure}
\includegraphics[angle=0, scale=0.7]{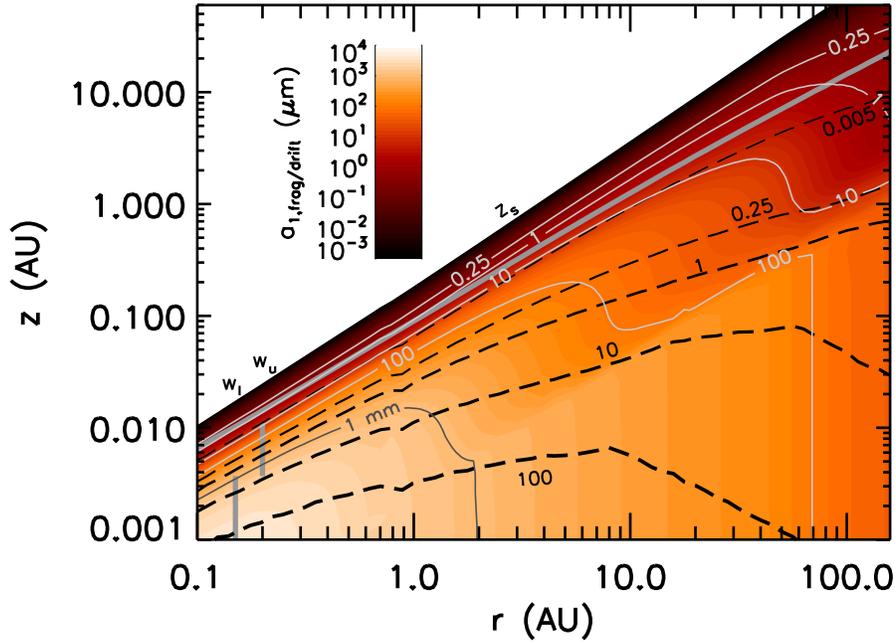}
\caption{V836 Tau: Analogous content with Fig. \ref{citau_frag}, and the same symbols. \label{v836tau_frag}}
\end{figure}

Since our sample has a range of accretion rates, with varying midplane temperatures due to viscous heating, we can compare the maximum grain sizes predicted by Eq. \ref{birnstiel8} using our best-fitting temperature- and density-structures to the $a_{max}$ found by our models.  The predicted maximum grain sizes (not accounting for settling) for CI Tau and V836 Tau, the highest and lowest accreters, respectively, are shown in Figs. \ref{citau_frag} and \ref{v836tau_frag}.  In these figures, we overplot the predicted heights above which grains of a given size should be depleted \citep[not accounting for fragmentation,][]{dd04} by settling to the midplane.  Additionally we show the location of the two-layer walls and the disk photospheres.  The maximum grain size in the disk implied at the location of CI Tau's lower wall layer ranges from 1 to 5$\mu$m, consistent with our observed $a_{max}$ of 3$\mu$m; for the upper wall layer it is less than 1$\mu$m, consistent with 0.25$\mu$m.  In both cases, the depletion heights for each grain size match closely the contours indicating the limits to where grains of that size can form from fragmentation theory.  In contrast, for the temperatures in the V836 Tau disk the grains can grow greater than 1mm in the inner 2 AU, which at first glance appears to be at odds with our observed results. However, the predicted depletion heights are lower by more than a factor of 10 than the maximum grain size contours, for a given grain size.  According to Figures \ref{citau_frag} and \ref{v836tau_frag}, the upper wall contains submicron grains, while the lower wall is dominated by grains too large to be visible to us via either silicate-feature emission or the bulk of the continuum emission.

In addition to depletion in the upper layers, we may see indirect evidence for dust-gas mass ratio enhancement in the lower layers of the wall by comparing the wall temperature, $T_{wall}$, with predicted silicate sublimation temperatures, $T_{sub}$.  If the silicate dust sublimates under equilibrium conditions, it can do so in one of two ways: in kinetic equilibrium or in chemical equilibrium.  Kinetic equilibrium is typically expected for environments in which sublimation is purely a thermal decomposition.  Chemical-equilibrium dust composition is expected at high densities for temperatures approaching that of sublimation \citep[e.g.][]{gail04}.

Chemical equilibrium is expected for environments in which the dust and gas can engage in reactions.  The chemical composition of the gas is of importance, particularly the oxygen content relative to hydrogen.  In low oxygen environments, $H_2$ gas can react with the oxygen atoms in the silicate grains to make water vapor, effectively `chemi-sputtering' the grains at temperatures $\sim$150 K lower than $T_{sub}$ in the kinetic equilibrium case \citep{gail+99, rietmeijer+11}.  Conditions that favor sublimation in kinetic equilibrium over chemical equilibrium include lower ambient pressure (fewer gas-grain encounters) or oxygen-rich gas (gas-grain reactions less favorable).  

We compare the derived wall temperatures with the temperatures and pressures of each disk in the right hand side of Fig. \ref{diskstructs}.  In the wall upper layers, we find $T_{wall}$ consistent with the $T_{sub}$ predicted by chemi-sputtering of silicates in chemical equilibrium at the pressure of the disk surface.  However, the lower layers have $T_{wall}$ more consistent with a hotter $T_{sub}$ expected from kinetic equilibrium sublimation at the pressure of the midplane.  Since the midplane has high densities, we had expected it to be in chemical equilibrium.  The fact that we see kinetic equilibrium $T_{sub}$, may suggest  that there is an increase in the oxygen content of the midplane gas due to an increase in the dust-gas mass ratio by a factor of 50 to 500 \citep{rietmeijer+11}.  Self-consistent settling models predict that the midplane dust-gas ratio can be enhanced from a value of $\sim$0.01 to at most 0.2 \citep{mulders+11}.  However, there are mechanisms, e.g. dust filtration, which can reduce the dust-gas ratio on the inward side of a change in the surface density  \citep[in the case of][a gap opened by a planet]{zhu+12}.  A change in surface density, and subsequent particle trapping, is also induced a the location in the disk where each type of dust sublimates, e.g. the snowline \citep{kretke+07}.  It may be possible to build up the dust-gas ratio in the lower layer of the wall, which is by definition the silicate `snowline', in this manner.

\subsection{Scenarios for spatial variation of silicate iron content and stoichiometry}

An intriguing result from the physical models is the suggestion of silicates of pyroxene stoichiometry with an enhanced iron content in the lower layer of the wall.  The best fits to the 2 to 5 $\mu$m region came from the more iron-rich silicates, with a fraction of iron between 30 and 50\%, or $x$=0.5$-$0.7.  At first, our result appears in contrast with other mineralogy studies of T Tauri disks.  Modeling of crystalline olivines in mid- to far-infrared spectra of gas-rich systems have found consistently high Mg/(Mg+Fe) fractions, e.g. $x$ $>$0.9 to $>$0.99 \citep[][respectively]{tielens+98,mulders+11}.  Analyses of more mature systems, i.e. debris disks, find mixed results that may be consistent with a radial dependence of the iron content; \citet{olofsson+12} find two `warm' debris disks with $x\sim0.2$ from fits to their IRS spectra, while \citet{devries+12} finds $x$=0.99$\pm$0.001 for the cool debris disk around $\beta$ Pictoris.  This predominance of crystalline Mg-rich olivines is consistent with differences in the sublimation and annealing temperatures of the two olivine end-members.  Specifically, Fe-rich silicates require a higher temperature than Mg-rich silicates to anneal \citep[1400 vs 1100 K, respectively, at an unspecified pressure,][]{nuthjohnson06}, while their stability limiting temperature against sublimation for a given pressure is lower than Mg-rich silicates \citep[e.g. 1225 vs. 1375 K, respectively, at 100 dyn cm$^{-2}$ assuming chemical equilibrium with an H$_2$ reservoir,][]{gail+99}.

However, the paucity of crystalline Fe-rich olivine in astrophysical observations does not necessarily imply that Fe-rich amorphous grains are also absent.  Using spectral decomposition models to fit the 10$\mu$m silicate complex in disks \citet{sargent+09} find that, independent of their stoichiometry, large amorphous silicates have equal Fe and Mg content ($x$=0.5).

Presolar silicates recovered from meteorites have roughly equal amounts of iron and magnesium, $x\sim$0.5, at micron scales.  Closer examination shows that these particles are comprised of individual submicron grains with either a wholly Fe-rich or Mg-rich composition \citep{paquette+11}, consistent with experiments by \citet{rietmeijer+99} for solids condensing from a Mg-Fe-SiO-H$_2$-O$_2$ gas.  Although iron is equally represented in presolar silicates, there is a radial gradient in the iron content of solar system silicates, from $x=0.7-0.8$ in the S-type asteroid 25143 Itokawa \citep[][$\sim$0.95 to 1.7 AU]{nakamura+11} to $x>$90\% in the Jupiter-family comet Wild-2 \citep[][$\sim$1.6 to 5.3 AU]{zolensky+06}.  

The question is then how to interpret the division in composition between the upper layer of the wall and the lower layer, both in terms of the iron content and the pyroxene stoichiometry.  If we have a 50-50 mix between Fe- and Mg-rich amorphous olivine in the upper layer of the disk, the Fe-rich olivine should sublimate preferentially, leaving the annealed forsterite (which is by definition Mg-rich) and some remainder of the mixed Mg-Fe olivine.  In tandem with sublimation, if the surrounding gas is oxygen poor (consistent with the agreement of our upper wall temperature-pressure combinations with the P-T relationship in chemical equilibrium), then olivine with a mixed Mg-Fe composition could destabilize into pyroxene-metal assemblages \citep{matas+00}.  Since the inner disk is turbulent, these end products would be mixed vertically; in the upper layers, Fe-rich grains could not survive, but in the midplane they might.  Additionally, dust in the optically thin portions of the wall would experience irradiation by ions from the star.  For initially crystalline olivine, this has the effect of changing it to an amorphous, glassy pyroxene stoichiometry, as demonstrated by \citet{rietmeijer09}, and references therein.  This author also finds a specific instance in which evidence for this reaction is recorded in a chondritic aggregate interplanetary dust particle, with the glassy pyroxene product having an Mg/(Mg+Fe)=0.74$\pm$0.1, consistent with our best-fitting lower wall composition.

\section{Conclusions}

We have combined a simple approximation of a curved dust sublimation wall with self-consistently calculated physical disk models to a) test if disk walls need to be curved, b) determine what their structure and dust content is, and c) compare these properties with those of the disk.  From this work, our main conclusions are:

\begin{itemize}

\item The 2-10$\mu$m excess in T Tauri stars is best fit by dust sublimation walls that are curved by the triple effects of the pressure structure, grain size distribution, and grain composition in the disk.  

\item Walls that fit the largest NIR excesses in our sample are not significantly elevated above the disk surface height, where the stellar radiation is absorbed, and therefore do not shadow the outer disk.  The decrease in the slope of the SED of these disks is attributable to dust depletion on the order of 0.01 to 0.0001 from the upper layers.

\item The grain size distribution in the wall may evolve as a function of the mass accretion rate, due to turbulent mixing.  Current grain growth theories that include fragmentation are consistent with our observations and predict $>$1 mm grain production in the inner 2 AU of the disk with the lowest \Mdot and $\alpha$.

\item Large, iron-rich pyroxene grains are required to fit the NIR excess.  The iron content required in the pyroxene is 40$^{+10}_{-20}$\% which, combined with recent {\it Herschel} studies reporting 10\% or less in forsterite beyond 10 AU, is suggestive of the iron gradient found in the solar system.  

\end{itemize}

We obviously cannot make any absolute conclusions regarding the wall dust content, given the limited number of compositions we tried compared with the large variety of minerals that could exist in the disk.  However, this study highlights the importance of the dust composition and its implications on the young terrestrial planet forming region and demonstrates the need for more detailed future models taking into account the inner disk mineralogy, gas-phase chemistry, wall geometry, and disk structure.

\acknowledgments
This work is based on observations made with the NASA Infrared Telescope Facility.  This material is based upon work supported by the National Science Foundation Graduate Student Research Fellowship under Grant No. DGE 0718128.  P. D. acknowledges support from PAPIIT UNAM.  N.C acknowledges support from NASA Origins grants NNX08AH94G.  C.E. was supported by a Sagan Exoplanet Fellowship from the National Aeronautics and Space Administration and administered by the NASA Exoplanet Science Institute (NExScI).  K. L. was supported by grant AST-0544588 from the National Science Foundation.  The Center for Exoplanets and Habitable Worlds is supported by the Pennsylvania State University, the Eberly College of Science, and the Pennsylvania Space Grant Consortium.  This publication made use of NASA's Astrophysics Data System Abstract Service as well as the SIMBAD database and Vizier catalog service, operated by the Centre de Données astronomiques de Strasbourg.

\clearpage

\begin{deluxetable}{lllll}
\tabletypesize{\small}   
\tablewidth{0pt}
\tablecaption{Best-fitting model parameters}
\tablehead{\colhead{Parameter} & \colhead{CI Tau} & \colhead{DE Tau} & \colhead{GO Tau} & \colhead{V836 Tau}}
\startdata
Star \\
\hline \\
$T_{eff}$ (K) 	  & 4060  & 3720  & 3850 & 4060  \\
$R_{*}$ (\Rsun) & 1.41           &  2.10  & 1.22 & 1.76   \\
$M_{*}$ (\Msun) & 0.8          & 0.48  & 0.59 & 0.76   \\
\Mdot (\Msun/yr) 	& 2.9$\times$10$^{-8}$ & 2.2$\times$10$^{-8}$  & 3.8$\times$10$^{-9}$ & 1.9$\times$10$^{-10}$  \\
\hline \\
Wall, lower \\
\hline \\
$T_{wall,1}$ (K) 		  & 1750     &  1650  & 1200 &  1300  \\
$a_{max}$ ($\mu$m) & 3.0        &  2.0  & 1.0 & 0.5   \\
sil. comp. & PyMg60 (100\%) &  PyMg60 (100\%)  & PyMg60 (100\%) & PyMg60 (100\%)   \\
$R_{wall,1}$ (AU)          & 0.12         &  0.11  & 0.13 & 0.15   \\
$h_{wall,1}$ = $z_{wall,1}$ (AU)           & 1.2$\times$10$^{-2}$ (3.5H)      &  1.4$\times$10$^{-2}$ (3.5H)  & 2.1$\times$10$^{-3}$ (0.6H) & 3.7$\times$10$^{-3}$ (1H)   \\
$z_{s,disk} (R_{wall})$ (AU) &	1.0$\times$10$^{-2}$	&	8.8$\times$10$^{-3}$	& 9.85$\times$10$^{-3}$ & 1.08$\times$10$^{-2}$ \\
\hline \\
Wall, upper \\
\hline \\
$T_{wall,2}$ (K)             & 1200      & 1000	  & 800 & 1000  \\
$a_{max}$ ($\mu$m) & 0.25         & 0.25  & 0.1 & 0.25  \\
$sil.$ $comp.$ & PyMg60 (50\%) & OliMg50 (30\%) & OliMg50 (30\%) &  OliMg50 (40\%)  \\
		&				&		Fo (70\%)	 &  Fo (70\%)	& Fo (60\%) \\
$R_{wall,2}$ (AU)          & 0.23         & 0.23  & 0.17 & 0.20  \\
$h_{wall,2}$ (AU)	& 2.5$\times$10$^{-2}$ (3.9H) &	1.2$\times$10$^{-2}$ (1.5H) &  1.1$\times$10$^{-2}$ (2.4H)  & 1.1$\times$10$^{-2}$ (2H) \\
$z_{wall,2}$ (AU)          & 3.7$\times$10$^{-2}$     &  2.6$\times$10$^{-2}$  & 1.3$\times$10$^{-2}$ & 1.5$\times$10$^{-2}$  \\
$z_{s,disk} (R_{2})$ (AU) &	2.4$\times$10$^{-2}$	&	2.5$\times$10$^{-2}$	& 1.5$\times$10$^{-2}$ & 1.4$\times$10$^{-2}$ \\
\hline \\
Disk \\
\hline \\
$i$ (\degr)                  & 55 $^b$    & 40  & 65 & 65   \\
$R_{d}$ (AU) 		& 100 $^b$ & 100  & 140 & 140  \\
$\alpha$ 			& 0.005 & 0.05  & 0.002 & 0.00008  \\
$\epsilon$  		& 0.005   & 0.008  & 0.01 & 0.0002  \\
$a_{max,s}$ ($\mu$m) & 0.75  &  0.25  & 3.0 & 1.0  \\
$a_{max,b}$ (mm)    & 1  & 1  & 1 & 1  \\
silicates & PyMg80 (100\%) &  PyMg80 (90\%)  & PyMg80 (100\%) & PyMg80 (90\%)  \\
   & ...  &  Fo (10\%)  & ... & Fo (10\%)  \\
 $M_{disk}$ (\Msun) & 6.8$\times$10$^{-2}$  & 2.7$\times$10$^{-3}$  & 2.8$\times$10$^{-2}$ &  2.9$\times$10$^{-2}$

\enddata
\tablecomments{
\\
$1$ Stellar parameters are from \citet{mcclure+13}. \\
$2$ In all cases, the distances to Taurus was taken to be 140 pc. \\
$^3$ This is the $T_{eff}$ of the star, used in the wall calculations.  As discussed in \S\ref{diskfits}, CI Tau has an accretion luminosity almost a large as its stellar luminosity.\\
{\it References}:  $^a$ \citet{kenyon+94}, $^b$ resolved at 880 $\mu$m by \citet{aw07}
}
\label{alldiskfits}
\end{deluxetable}
   
\end{document}